\documentclass[reprint,raggedbottom,aps,prmaterials,amsfonts,amsmath,amssymb,floatfix,superscriptaddress]{revtex4-2}

\usepackage{graphicx,hyperref} 
\usepackage[all]{hypcap} 
\hypersetup{colorlinks=true,linkcolor=blue,citecolor=blue,urlcolor=blue}

\newcommand{\Ang}[0]{\,\mathring{\mathrm{A}}} 

\renewcommand{\v}[1]{\mathbf{#1}}
\newcommand{\uv}[1]{\mathbf{\hat{#1}}}
\newcommand{\gv}[1]{\boldsymbol{#1}}

\begin{document}

\title{A data-centric framework for crystal structure identification in atomistic simulations using machine learning}

\author{Heejung Chung} 
\thanks{These authors contributed equally to this work}
\affiliation{Department of Computer Science, Stanford University, Stanford, California 94305, USA}
\affiliation{Department of Materials Science and Engineering, Stanford University, Stanford, California 94305, USA}
\affiliation{Cavendish Laboratory, Department of Physics, University of Cambridge, J. J. Thomson Avenue, Cambridge CB3 0HE, United Kingdom}

\author{Rodrigo Freitas}
\thanks{These authors contributed equally to this work}
\affiliation{Department of Materials Science and Engineering, Stanford University, Stanford, California 94305, USA}
\affiliation{Department of Materials Science and Engineering, Massachusetts Institute of Technology, Massachusetts 02139, USA}

\author{Gowoon Cheon}
\altaffiliation{Current affiliation: Google Research, Mountain View, California 94043, USA}
\affiliation{Department of Applied Physics, Stanford University, Stanford, California 94305, USA}

\author{Evan J. Reed}
\thanks{\href{https://news.stanford.edu/report/2022/03/28/evan-reed-computational-materials-scientist-used-ai-discover-ideal-engineering-materials-died/}{Dedicated in memory of Evan J. Reed.}}
\affiliation{Department of Materials Science and Engineering, Stanford University, Stanford, California 94305, USA}

\date{\today}
\begin{abstract}
  Atomic-level modeling performed at large scales enables the investigation of mesoscale materials properties with atom-by-atom resolution. The spatial complexity of such cross-scale simulations renders them unsuitable for simple human visual inspection. Instead, specialized structure characterization techniques are required to aid interpretation. These have historically been challenging to construct, requiring significant intuition and effort. Here we propose an alternative framework for a fundamental structural characterization task: classifying atoms according to the crystal structure to which they belong. Our approach is data-centric and favors the employment of Machine Learning over heuristic rules of classification. A group of data-science tools and simple local descriptors of atomic structure are employed together with an efficient synthetic training set. We also introduce the first standard and publicly available benchmark data set for evaluation of algorithms for crystal-structure classification. It is demonstrated that our data-centric framework outperforms all of the most popular heuristic methods -- especially at high temperatures when lattices are the most distorted -- while introducing a systematic route for generalization to new crystal structures. Moreover, through the use of outlier detection algorithms our approach is capable of discerning between amorphous atomic motifs (i.e., noncrystalline phases) and unknown crystal structures, making it uniquely suited for exploratory materials synthesis simulations.
\end{abstract}

\maketitle

\section{Introduction}
  Atomic-level computational modeling enables the calculation of materials' properties while taking into account the contribution of each constituent atom individually. When such simulations are large enough ($\gtrsim 10^6$ atoms) -- also known as cross-scale -- they enable the understanding of mesoscale materials phenomena with atomic resolution \cite{luis_natmat,freitas_2020,luis_nature,curtin_1,stishovite,jaime}. Nevertheless, the spatial complexity of cross-scale simulations renders them unsuitable for simple human visual inspection, severely limiting the amount and quality of scientific information that can be extracted. Widespread access to computing capabilities that were not long ago available only to select research centers has heightened the need for algorithms that aid and augment human intuition in interpreting large-scale atomistic simulations \cite{stuko_1,dxa,nussinov}.

\begin{figure*}[tb]
    \centering
    \includegraphics[width=\textwidth]{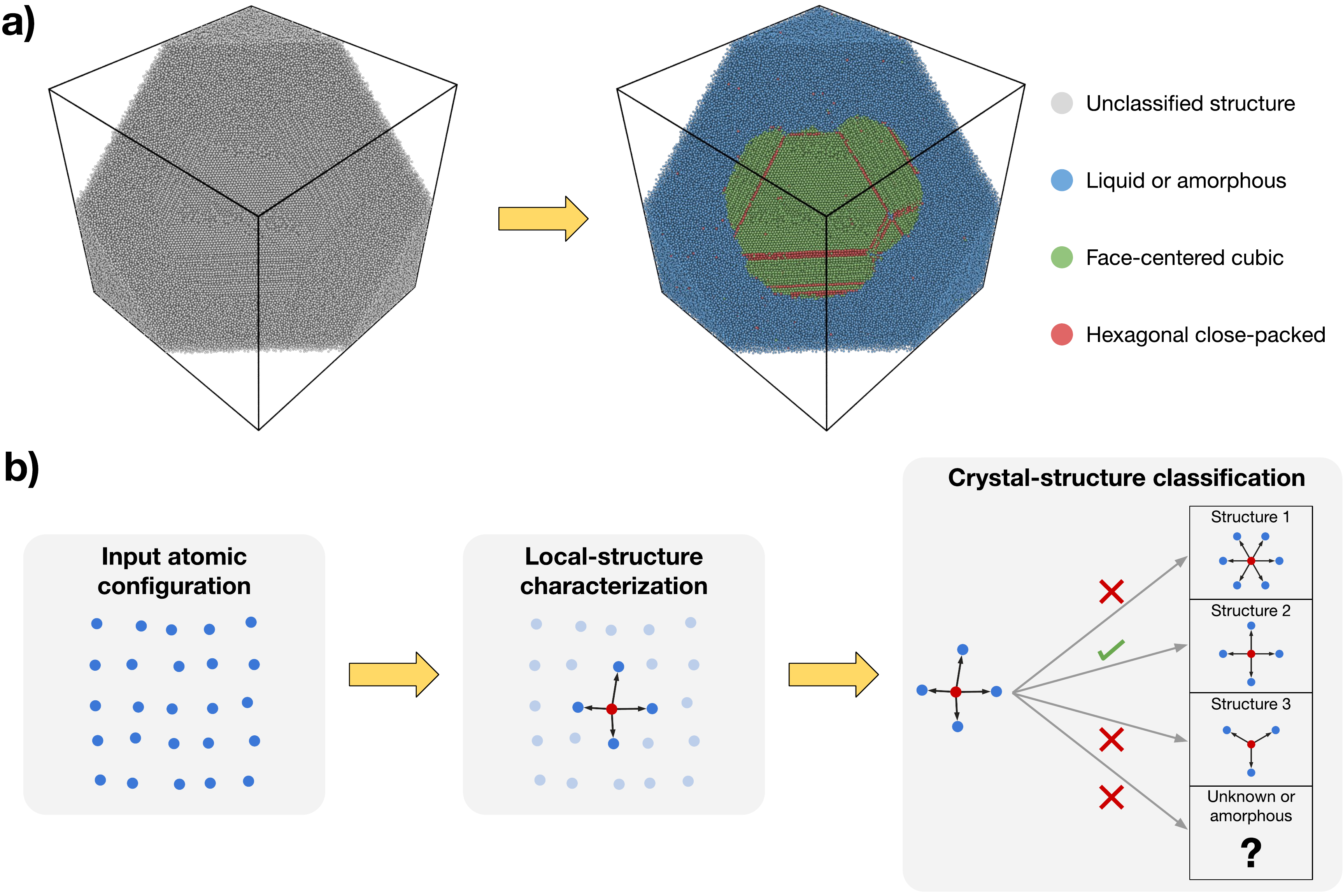}
    \caption{\label{fig:crystal_structure_classification} Illustration of the fundamental tasks performed by a crystal-structure classification algorithm. \textbf{a)} Crystal-structure classification of a Molecular Dynamics simulations of copper crystallization with two million atoms. On the left the input atomic configuration of a cross section of a simulation snapshot is shown. On the right each atom has been color coded according to its local structure. Notice how it becomes easy to visually discern the growing crystal from the melt. Moreover, it is also possible to identify microstructural elements of the growing crystal: the crystal structure is face-centered cubic while stacking-fault defects are easily identified by atoms with local hexagonal close-packed structure. \textbf{b)} The starting point of a crystal-structure classification algorithm is the atomic configuration provided by an atomistic simulation snapshot. One of the atoms from the simulation is selected and its local structure is characterized by quantifying a set of structural features that capture metrics of the local symmetry. During the crystal-structure classification step the structural features are employed to make a decision about which crystal structure in the algorithm database best approximates the atom's local structure. It is possible that the local structure is unknown or even amorphous (such as in liquids and glasses), thus the classification algorithm should be able to detect such cases.}
  \end{figure*}

  The ability to detect ordered atomic motifs lies at the heart of most algorithms designed to assist in the interpretation of large-scale simulations. This local crystal structure identification is important because the structure surrounding atoms constituting microstructural elements (i.e., crystal defects) is different from the structure of crystalline lattices (see Fig.~\ref{fig:crystal_structure_classification}a).

  There exists a myriad of crystal-structure classification methods \cite{ptm,cna_1,cna_2,acna,icna,vorotop,chillp,aja} employing a variety of different approaches such as computational graphs, advanced geometrical algorithms, and sometimes relying solely on sets of ad hoc empirical rules \cite{aja}. Yet, despite such variety, most crystal classification methods share similar drawbacks. For example, none of the available methods can be systematically generalized to new structures or to systems with an arbitrary number of chemical elements, making them unsuitable for high-throughput approaches \cite{aflow,high_throughput,mp,fireworks} that are pervasive nowadays in materials science. While expert domain knowledge is useful in developing heuristics that work well for particular tasks, there is no guarantee of out-of-distribution generalization to different systems. Another marked lapse in the field is the lack of a rigorous benchmark comparison of the performance of each method, despite their importance and widespread application. We also note that none of the currently available methods can discern between disordered atomic motifs -- such as present in liquids and glasses -- and unknown crystal structures, making them unsuitable for exploratory materials synthesis simulations \cite{stishovite,freitas_2020}.

  Recently, numerous Machine Learning (ML) approaches have been developed with the capacity to process information contained in atomistic simulations, see for example Refs.~\cite{ml_struct_1,ml_struct_2,ml_struct_3,ml_struct_4,ml_struct_5,nussinov,freitas_2020,struct_1,struct_2,struct_3,struct_4,struct_5}. These efforts have unequivocally established the ability of ML algorithms to find subtle but meaningful patterns in the structure and dynamics of atomic motion. The applications of these new algorithms have been various, ranging from the structural analysis of complex atomic arrangements at a various scales \cite{ml_struct_1,ml_struct_2,ml_struct_3,nussinov} to the creation of novel physical models about the behavior of materials \cite{ml_struct_4,ml_struct_5,freitas_2020}. 
  
  In this article we leverage ML and other data-science tools in order to develop a complete framework for structure classification that functions similarly to well-established and widely-adopted crystal-structure classification methods \cite{ptm,cna_1,cna_2,acna,icna,vorotop,chillp,aja}. Unlike existing heuristic methods, our framework for crystal-structure classification, namely the data-centric crystal classifier (which we will refer to as DC3), does not rely on predefined programmed rules created by the intuition of human experts. This makes DC3 easily extendable to chemically complex systems, for which heuristics would be difficult to construct. Through a careful statistical analysis we demonstrate that this data-centric approach has better accuracy than any of the most popular heuristic algorithms. Moreover, through a variety of examples we establish the capacity of this data-centric framework to be systematically generalized to different crystal structures and systems with arbitrary chemical complexity (i.e. structures with multiple chemical elements), while remaining highly transferable (i.e. material independent), and robust against thermal noise.
  \begin{figure*}[tb]
      \centering
      \includegraphics[width=\textwidth]{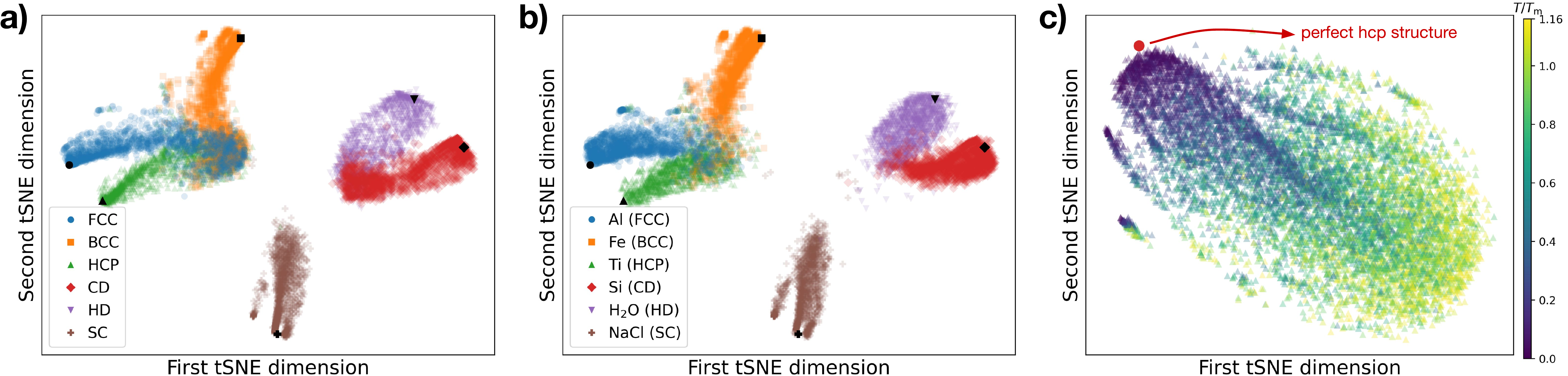}
      \caption{Distribution of different crystal structures in the high-dimensional space of feature vectors as visualized through tSNE (see Appendix \ref{app:tSNE} for details on the application of tSNE). \textbf{a)} tSNE results for a randomly selected subset of the synthetic data set, demonstrating that the local-structure characterization approach developed here can effectively distinguish between crystal structures. Black data points correspond to the feature vector of perfect crystal structures (i.e., undisturbed by thermal noise). \textbf{b)} Data from Molecular Dynamics simulations (with temperatures ranging from $0.04\,T_\text{m}$ to $1.32\,T_\text{m}$) have a similar tSNE distribution as the data from the synthetic data set, demonstrating the adequacy of the synthetic data set for training the Machine Learning model. \textbf{c)} tSNE plot of the temperature-resolved distribution of the hcp structure feature vectors. Data points corresponding to high temperatures (i.e., large thermal noise) are the most challenging cases to classify due to their increased distance from the perfect reference structure (shown as the red circle for the hcp structure).}
      \label{fig:tsne}
  \end{figure*}

\section{Methods}
  \subsection{Local-structure characterization}
  The fundamental task of a crystal classification algorithm is to assign a label $\tilde{y}_i$ to each atom $i$ according to its surrounding structure (Fig.~\ref{fig:crystal_structure_classification}b), where each label is uniquely associated with a crystal structure. Heuristic algorithms for crystal classification drawn from the knowledge of experts about crystal symmetries in order to make such classification decisions. Whichever crystal-structure features an expert considers important is converted into an algorithm that, given an atom $i$, quantifies such features and makes a decision about the most appropriate crystal structure to be assigned.

  Our approach does not rely on expert knowledge to select features that may be important. Instead, the structure surrounding atoms is characterized as completely as possible using a variety of symmetry and structure functions. The task of assigning a crystal structure based on such characterization is then delegated to an ML algorithm. Hence, the final format of this local-structure characterization must be appropriate to be used with ML algorithms, which means that each atom $i$ must have an associated high-dimensional feature vector $\mathbf{x}_i$ describing its local structure.

  Each feature vector $\mathbf{x}_i$, capturing the local structure of atom $i$, was constructed using two families of functions. The first family of functions employed is known as Steinhardt parameters \cite{steinhardt}
  \begin{equation}
    \label{eq:steinhardt}
    Q_\ell^{N_\text{b}}(i) = \sqrt{\frac{4\pi}{2\ell+1} \sum_{m=-\ell}^\ell \left| q_{\ell, m}^{N_\text{b}}(i)\right|^2} ,
  \end{equation}
  where $i$ is the number of the atom for which the local neighborhood is being described, $N_\text{b}$ is the number of nearest neighbors being considered, and $\ell$ is an integer. The quantity $q_{\ell, m}^{N_\text{b}}(i)$ is defined as
  \begin{equation}
    \label{eq:q_lm}
    q_{\ell, m}^{N_\text{b}}(i) = \frac{1}{N_\text{b}} \sum_{j=1}^{N_\text{b}} Y_\ell^m(\uv{r}_{ij})
  \end{equation}
  where $Y_\ell^m(\uv{r}_{ij})$ are spherical harmonics and $\uv{r}_{ij}$ is the unit vector along the direction connecting atom $i$ to atom $j$. Steinhardt parameters are functions designed to capture purely angular information about the local structure (see Eqs.~\eqref{eq:steinhardt} and~\eqref{eq:q_lm}). More specifically, for each value of $\ell$ this function is sensitive to different orientational symmetries present in the local-neighborhood of atom $i$ as defined by its first $N_\text{b}$ neighbors. The second family of functions is known as radial structure functions (RSF) \cite{rsf}: 
  \begin{equation}
    \label{eq:rsf}
    G_{r,\sigma}^{N_\text{b}}(i) = \sum_{j=1}^{N(r_\text{cut})} \exp\left[ -(r_{ij} - r)^2/2\sigma^2 \right]
  \end{equation}
  where $N(r_\text{cut})$ is the number of neighbors of atom $i$ within cutoff radius $r_\text{cut}$, $r$ and $\sigma$ are free parameters, and $r_{ij}$ is the distance between atoms $i$ and $j$. The RSFs are designed to capture purely radial information about the local structure (see Eq.~\eqref{eq:rsf}). More specifically, this function quantifies the density of atoms at distance $r$ from atom $i$ with a spatial resolution controlled by $\sigma$. The feature vectors $\mathbf{x}_i$ are composed of both families of functions, RSF and Steinhardt parameters, with each vector component corresponding to one of these functions evaluated using a set of parameters: ($\ell, N_\text{b}$) for Steinhart features or ($r, \sigma, N_\text{b}$) for RSFs (see the Appendix \ref{app:feature_vector} for details on the selection of the numerical values of these parameters). Notice that the feature vector defined above does not account for chemical complexity (i.e., all atoms are considered to belong to the same species). In Sec.~\ref{sec:chemically_complex} we demonstrate how for structures with multiple chemical elements can be accounted for in this data-centric framework.

  The local-structure characterization process needs to generalize across different materials sharing the same crystal structure (e.g., Al and Cu both have face-centered cubic structures but the lattice constant is $4.1\Ang$ for Al and $3.6\Ang$ for Cu). Thus, a discussion about RSF parameters $r$ and $\sigma$ is warranted. In order to render $\v{x}_i$ independent of the material lattice constant the parameters $r$ and $\sigma$ are not preselected before the local-structure characterization takes place. Instead, these parameters are determined with regard to a local-distance metric calculated only during the evaluation of $\v{x}_i$, which effectively renders $\v{x}_i$ independent of the magnitude of the lattice constant (see the Appendix \ref{app:feature_vector} section for more details on the numerical algorithm to evaluate the local-distance metric). This approach was inspired by the adaptive cutoff method developed by \citet{stuko_adaptive}.

  \subsection{Unbiased data generation for training} An ML algorithm learns how to assign crystal structure labels $\tilde{y}_i$ from local-structure feature vectors $\mathbf{x}_i$ by observing patterns in a data set: $\left\{(y_i,\mathbf{x}_i)\right\}$, where $(y_i,\mathbf{x}_i)$ are pairs of correctly associated labels and feature vectors. Due to crystal structure distortions caused by thermal noise many different feature vectors $\mathbf{x}_i$ are associated with the same label $y_i$. Hence, it is important that the training of an ML model is performed on a data set consisting of examples drawn from structure distortions corresponding to a variety of thermal noise amplitudes.

  In order for DC3 to generalize across different materials sharing the same underlying crystal structure the training data set must not be associated with any particular material chemistry. This is because even materials with the same crystal structure at the same level of thermal noise have different vibration patterns (i.e., phonon spectrum) that generate different structure distortions. In order to not bias our ML model towards one preferred vibration pattern we introduce a data set created using random crystal structure distortions. This synthetic data set is built by first creating an undisturbed crystal structure where atoms lie exactly at their ideal positions. Then each atom is randomly displaced such that the displacements are on average uniformly distributed over a sphere of radius defined by the amplitude of thermal noise we desire to sample. The displacement amplitude is made with respect to the distance $d$ to the atom's first neighbor in the undisturbed crystal structure, with displacement radii as large as $0.25d$ being employed (see the Appendix \ref{app:synthetic} for a more detailed description of the algorithm).
 
  Synthetic data sets were built for six crystal structures: face-centered cubic (fcc), body-centered cubic (bcc), hexagonal close-packed (hcp), cubic diamond (cd), hexagonal diamond (hd), and simple cubic (sc). Three of these structures are fundamental Bravais lattices: fcc, bcc, and sc, while the other three are constructed from Bravais lattices by the addition of a basis (i.e., an atomic motif). A feature vector $\v{x}_i$ was computed for each atom in the synthetic data set. Visualization of the feature vectors distribution in their high-dimensional space can be performed through the application of a dimensionality reduction technique known as tSNE \cite{tsne}, as shown in Fig.~\ref{fig:tsne}a. This figure establishes that our approach for local-structure characterization can effectively distinguish between crystal structures, despite the presence of large distortions due to thermal noise.

  \subsection{Machine Learning classifier algorithm} 
\begin{figure}[tb]
      \centering
      \includegraphics[width=0.45\textwidth]{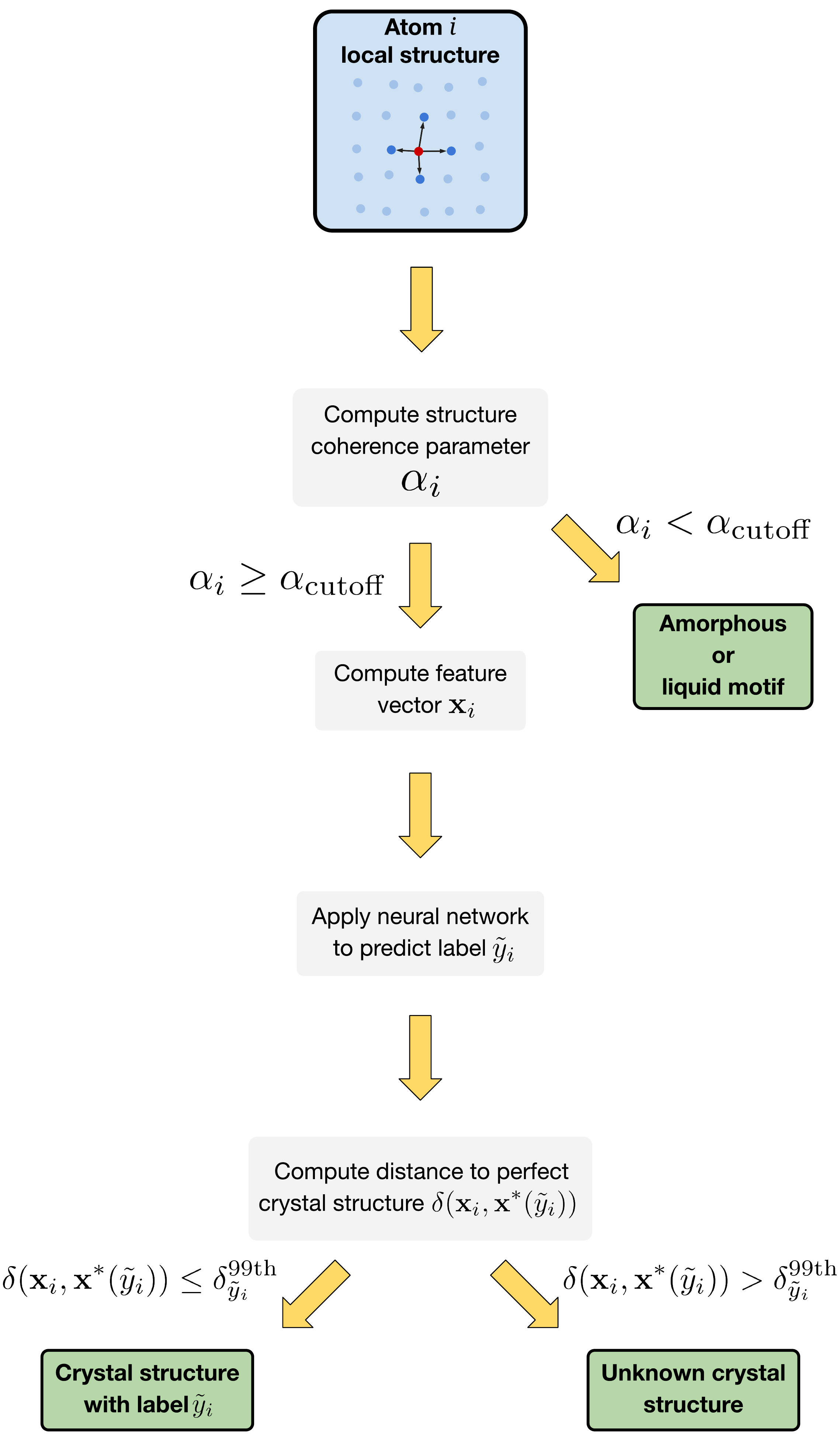}
      \caption{Data-centric crystal classifier (DC3) framework. The initial point of the framework is the input of the local structure surrounding atom $i$ (blue box). The algorithm ends when a label has been assigned to this local structure (green boxes). The feature vector of the perfect crystal structure (i.e., no thermal noise) associated with label $\tilde{y}_i$ is $\v{x}^*(y_i)$.}
      \label{fig:framework}
  \end{figure}

The next step in our data-centric framework is the introduction of a proper ML algorithm to predict a labels $\tilde{y}_i$ for given feature vectors $\v{x}_i$. For this task we employed a multi-class feed-forward neural network composed of three hidden layers with $100$ rectified linear units and softmax output. Training was performed using only data from the synthetic data set. See Appendix \ref{app:training} for more details on the neural network training and model selection. The neural network classification task is shown in the central part of the DC3 framework illustrated in Fig.~\ref{fig:framework}.
  
  \subsection{Recognizing amorphous motifs and unknown crystal structures} An important capability of any crystal-structure classification algorithm is its capacity to recognize amorphous motifs -- present in liquids and glasses -- and unknown crystal structures as such. In the absence of this capability a known crystal structure label $\tilde{y}_i$ would be mistakenly assigned, leading to false-positive errors. Hence, in order to discern amorphous motifs from unknown crystal structures we have developed an outlier detection algorithm suitable to our data-centric framework. Next we describe the outlier detection algorithm and how it fits in the DC3 framework, illustrated in Fig.~\ref{fig:framework}.

  Given the local structure surrounding atom $i$, the first step of the outlier detection procedure is to determine whether this local structure corresponds to a crystalline motif or an amorphous one. Physically, the difference between these two motifs is that if atom $i$ is in a crystalline motif there is a high degree of similarity between its local structure and the local structure of its neighbors, while such similarity does not exist for amorphous motifs. Thus, we define the structure coherence parameter $\alpha_i = (1/N_\text{b}) \sum_{j=1}^{N_\text{b}} \gv{\xi}_i \cdot \gv{\xi}_j$ between $i$ and its first $N_\text{b}$ neighbors, where $\gv{\xi}_i$ is a vector composed of a combination of spherical harmonics in such way that the closer the structure surrounding atom $i$ and $j$ is the more parallel vectors $\gv{\xi}_i$ and $\gv{\xi}_j$ become. If $\alpha_i > \alpha_\text{cutoff}$ the local motif surrounding atom $i$ is considered crystalline, otherwise it is considered amorphous (our definition of the coherence parameter $\alpha_i$ was inspired by the study of \citet{alpha_frenkel}). The values of $\alpha_\text{cutoff}$ and $N_\text{b}$ were determined using only the synthetic data sets, see Appendix \ref{app:outlier} for more details on vector $\gv{\xi}_i$ and the coherence parameter $\alpha_i$.
  
\begin{figure}[tb]
      \centering
      \includegraphics[width=0.45\textwidth]{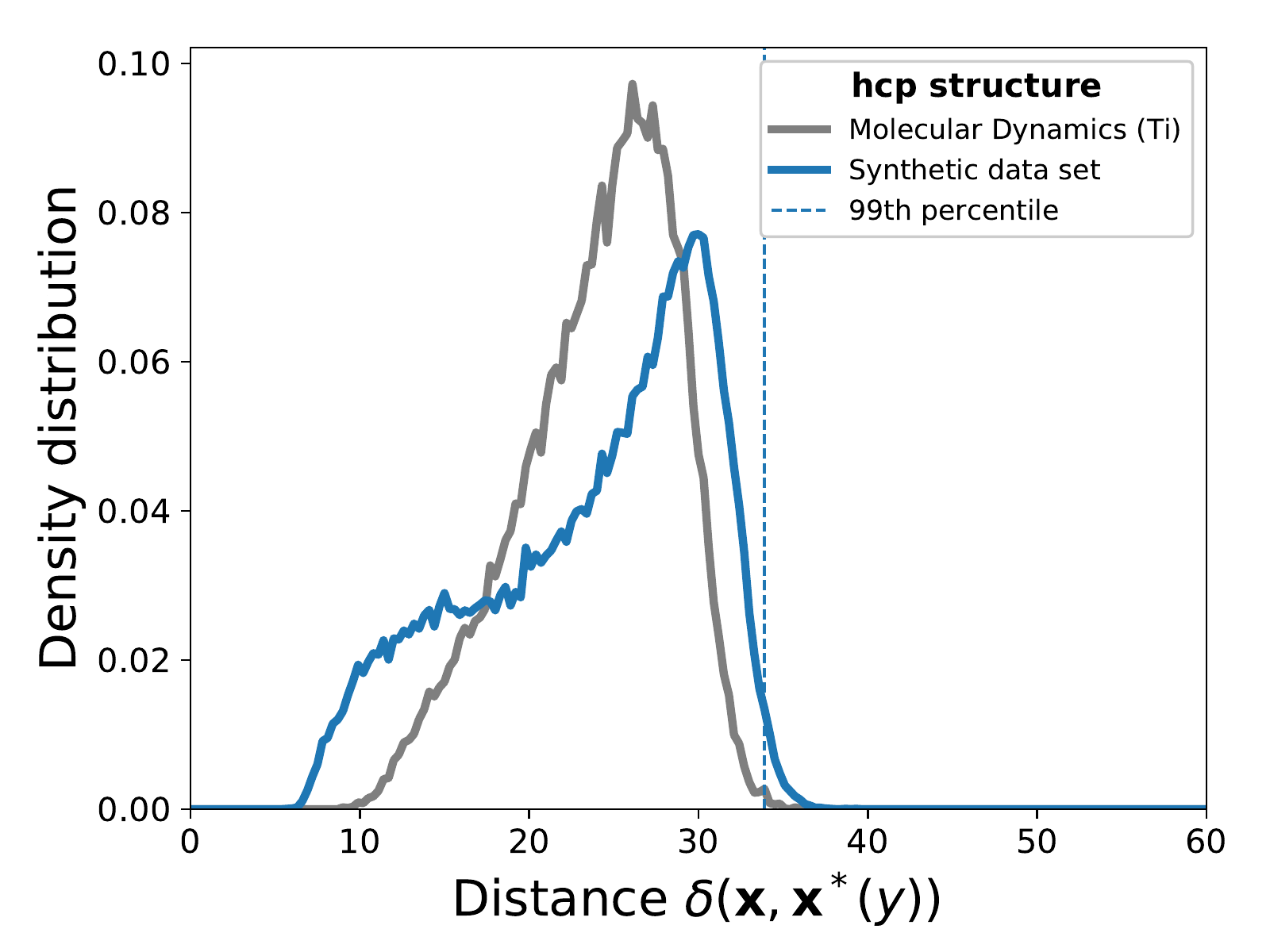}
      \caption{Histogram of distances of the feature vectors to the point corresponding to the perfect hcp structure, i.e. $\delta(\v{x},\v{x}^*(y))$ with $y = \text{hcp}$. The Molecular Dynamics data set (with temperatures ranging from $0.04\,T_\text{m}$ to $1.16\,T_\text{m}$) is depicted in gray while the synthetic data set is shown in blue. The 99th percentile of distances $\delta_{y}^{99\text{th}}$ for the synthetic data set is marked as the dashed blue line.}
      \label{fig:distance_density}
  \end{figure}

  Continuing with the example of atom $i$, let us assume that the local motif was identified as crystalline (i.e., $\alpha_i > \alpha_\text{cutoff}$). In this case, our framework must determine next whether the local structure surrounding atom $i$ corresponds to crystalline motif seen during training or to an unknown crystal motif. First the feature vector $\v{x}_i$ is evaluated for atom $i$ and the neural network trained previously is employed to predict a label $\tilde{y}_i$. Once this has been done the outlier detection procedure must determine if the assigned label $\tilde{y}_i$ is appropriate or if, instead, $\v{x}_i$ corresponds to an unknown crystal structure. In order to make that decision notice that each crystal structure has a feature vector that is special compared to all others: the one corresponding to the undistorted crystal structure (i.e., no thermal noise), shown in Fig.~\ref{fig:tsne}. For a crystal structure with label $y_i$ this special feature vector is $\v{x}^*(y_i)$. Hence, the distance $\delta(\v{x}_i,\v{x}^*(\tilde{y}_i))$ from $\v{x}_i$ to $\v{x}^*(\tilde{y}_i)$ is evaluated and compared to the location of the $99$th percentile ($\delta_{\tilde{y}_i}^{99\text{th}}$) of the distance density distribution as obtained from the synthetic data set (see Fig.~\ref{fig:distance_density} for an example for the case of the hcp structure). If $\delta(\v{x}_i,\v{x}^*(\tilde{y}_i)) \le \delta_{\tilde{y}_i}^{99\text{th}}$ the predicted label $\tilde{y}_i$ is considered appropriate, otherwise $\v{x}_i$ is determined to belong to an unknown crystal structure.

\section{Results}
  \subsection{Benchmark comparison against heuristic methods.} Comparison of performance between different crystal-structure classification algorithms should ideally be done on standardized data sets consisting of relevant and realistic materials simulations designed to push the limits of the methods capabilities. Such standard does not currently exist, which led us to develop and make publicly available the following benchmark data set.

  One or more representative materials and corresponding interatomic potentials were chosen for each of the six crystal structures considered in this article (see Table \ref{tab:materials_and_potentials}). The family of materials considered include metals, semiconductors, ionic solids, and molecular crystals. The choice of potentials span a variety of the most employed potentials in atomistic simulations, ranging from simple analytic potentials to state-of-the-art ML potentials. Molecular Dynamics (MD) simulations (described in Appendix \ref{app:MD}) with approximately $17000$ atoms were performed for each material at their respective melting temperatures at zero pressure. Over the course of the simulations configuration snapshots were saved at time intervals long enough to guarantee statistical independence between the snapshots. The accuracy of crystal-structure classification methods is then to be evaluated in each set of snapshots. Employing $T_\text{m}$ as the simulation temperature has the benefit of resulting in similarly large levels of the thermal noise for all materials, while being a temperature accessible to all crystal phases. Yet another MD simulation was performed for each potential in the liquid phase at $T = 1.6 \, T_\text{m}$. This group of simulations is to be employed in verifying the capability of crystal-structure classification methods to recognize amorphous atomic motifs and avoid false-positive crystal labels.
  \begin{table*}[tb]
    \centering
    \begin{tabular*}{0.8\textwidth}{c | @{\extracolsep{\fill}} c c c c c}
      \hline \hline
                & Structure & Material & Potential & $T_\text{m}$ [K] & $T_\text{ml}$ [$T_\text{m}$] \\
                \hline
                & fcc & Al & EAM \cite{eam_Al} & 933 & 1.16 \\
                & fcc & Ar & Lennard-Jones \cite{lj_Ar} & 83.8 & 1.12 \\
                & bcc & Li & SNAP \cite{snap_Li} & 454 & 1.20 \\
                & bcc & Fe & EAM \cite{eam_Fe} & 1811 & 1.08 \\
      benchmark & hcp & Ti & MEAM \cite{meam_Ti} & 1941 & 1.16 \\
       data set & hcp & Mg & EAM \cite{eam_Mg} & 923 & 1.16 \\
                & cd  & Si & EDIP \cite{edip} & 1687 & 1.16 \\
                & cd  & Ge & Tersoff \cite{tersoff_Ge} & 1211 & 1.32 \\
                & hd  & H$_2$O\footnote{The potential for H$_2$O is a coarse-grained monoatomic model where each molecule is represented by a single particle, resulting in the hexagonal diamond structure.} & Stillinger-Weber \cite{sw_mW}  & 273 & 1.32 \\
                & sc  & NaCl\footnote{In the benchmark data set both atom types (Na and Cl) were considered to be identical for the purpose of the crystal-structure classification, resulting in the simple cubic structure.} & Tosi-Fumi \cite{tosi_fumi_1,tosi_fumi_2} & 1074 & 1.16 \\
      
      \hline
                         & L1$_0$ & CuNi & EAM \cite{eam_CuNi} & 1462 & 1.24 \\
                         & L1$_2$ & CuNi & EAM \cite{eam_CuNi} & 1462 & 1.24 \\
      chemically complex & B1 (rock salt)  & NaCl\footnote{Differently from the benchmark data set, in the chemically complex data set both types of atoms (Na and Cl) are considered to be distinct.} & Tosi-Fumi \cite{tosi_fumi_1,tosi_fumi_2} & 1074 & 1.16 \\
      data set           & B2  & CoAl & EAM \cite{eam_CoAl} & 1913 & 1.04 \\
                         & B3 (zincblende) & InP & SNAP \cite{snap_InP} & 1355 & 1.20 \\
                         & B4 (wurtzite) & ZnO & ReaxFF \cite{reaxff_ZnO} & 850\footnote{Notice that the melting point for the ZnO potential is markedly different from the experimental melting point (2249K).} & 1.16 \\
      \hline \hline
    \end{tabular*}
    \caption{\label{tab:materials_and_potentials} Materials and interatomic potentials chosen to build data sets for the benchmark of crystal classification methods and the extension of DC3 to chemically complex systems (i.e. binary compounds). $T_\text{m}$ is the melting temperature and $T_\text{ml}$ is the highest temperature at which the material could be equilibrated in the crystal phase (i.e., the metastability limit). See Appendix \ref{app:MD} for details on the calculation of $T_\text{ml}$.}
  \end{table*}

  Next, we evaluate the performance of DC3 and other heuristic crystal classification methods in this independent and standardized benchmark data set. The following heuristic crystal classification algorithms were considered: Polyhedral Template Matching \cite{ptm} (PTM), Common-Neighbor Analysis \cite{cna_1,cna_2,acna} (CNA), interval CNA \cite{icna} (iCNA), Ackland-Jones analysis \cite{aja} (AJA), VoroTop \cite{vorotop}, and Chill+ \cite{chillp}. See Appendix \ref{app:benchmark} for the parameter selection and application details for each of these methods. As it can be seen on Table \ref{tab:accuracies} the DC3 method has the best accuracy for almost all materials considered, with the exception of Si and water where PTM and Chill$+$ are superior. Note, however, that in these two cases the accuracy of DC3 is already very high (above $99\%$) and only slighly inferior to other methods by about $0.8\%$. This is in contrast to cases such as Ti, where DC3 accuracy ($89.4\%$) is larger than the second best method by almost $7\%$. Another relevant observation is that DC3 maintains similar levels of accuracy across all six crystal structures, with the largest difference being a factor of $1.17$ between Fe (bcc) and Ge (cd). A similar analysis for heuristic methods results in factors of $1.21$ for PTM (Ti hcp vs water hd), $2.17$ for AJA (Li bcc vs Ar fcc), $2.65$ for VoroTop (Ar fcc vs Fe bcc), and $2.71$ for iCNA (Ti hcp vs Ar fcc). This suggests that expert knowledge biases algorithms towards better performance on structures for which the expert has better familiarity and intuition.
  \begin{table*}[tb]
    \centering
    \begin{tabular*}{\textwidth}{c | @{\extracolsep{\fill}} c c c c c c c c c c c}
      \hline \hline
        & Al (fcc)           & Fe (bcc)           & Ti (hcp)           & Si (cd)            & H$_2$O (hd)        & NaCl (sc)          & Ar (fcc)           & Li (bcc)           & Mg (hcp)           & Ge (cd)            & liquid \\ \hline
DC3     & \textbf{ 96.9 (3)} & \textbf{ 86.8 (5)} & \textbf{ 89.4 (5)} &          99.0 (1)  &          99.2 (1)  & \textbf{ 95.6 (3)} & \textbf{ 97.5 (2)} & \textbf{ 85.8 (5)} & \textbf{ 97.4 (2)} & \textbf{100.0 (0)} & 96.4 (1) \\
PTM\footnote{The parameters for PTM and Chill+ were highly optimized for each individual material (at $T = T_\text{m}$) using data from the same simulations employed to compute the reported accuracies.} &           95.9 (3) &           84.3 (5) &           82.8 (5) & \textbf{ 99.9 (1)} & \textbf{100.0 (0)} &           94.6 (3) &           96.9 (2) &           83.1 (5) &           95.7 (3) &           99.2 (1) & 99.5 (1) \\
iCNA    &           68.5 (6) &           56.6 (7) &           27.9 (6) &                 -- &                 -- &                 -- &           75.7 (6) &           51.6 (7) &           64.9 (6) &                 -- & 99.1 (1) \\
CNA     &           50.3 (8) &           39.9 (6) &           15.7 (5) &           98.8 (2) &           97.6 (2) &                 -- &           57.1 (7) &           34.4 (6) &           47.3 (6) & \textbf{100.0 (0)} & 100.0 (0) \\
AJA     &           66.9 (7) &           35.6 (7) &           42.4 (6) &                 -- &                 -- &                 -- &           74.0 (5) &           34.1 (7) &           67.0 (6) &                 -- & 84.3 (2) \\
VoroTop &           24.0 (6) &           61.2 (6) &           57.5 (7) &                 -- &                 -- &                 -- &           23.1 (6) &           57.2 (6) &           57.4 (6) &                 -- & 77.4 (2) \\
Chill+\footnotemark[1]  &                 -- &                 -- &                 -- & \textbf{ 99.8 (1)} &           98.8 (1) &                 -- &                 -- &                 -- &                 -- & \textbf{100.0 (0)} & 91.9 (2) \\
      \hline \hline
    \end{tabular*}
    \caption{\label{tab:accuracies} Accuracy comparison between the DC3 framework and the most popular heuristic algorithms as evaluated on the benchmark data set of Table \ref{tab:materials_and_potentials} (consisting of MD simulations at $T = T_\text{m}$). The largest accuracy for each crystal structure is highlighted in a bold font, while the $95\%$ confidence interval around the mean is shown in parentheses (see Appendix \ref{app:error} for details on the confidence interval calculation). The missing entries on the table are due to the fact that not all heuristic methods are capable of identifying all six structures. Notice that results for PTM and Chill+ were obtained using a set of highly optimized parameters for each individual material and temperature.}
  \end{table*}
  
  The best-performing heuristic method is PTM. Because of its excellent performance a discussion about some aspects of this method is warranted. In order to be used optimally PTM must first be fine-tuned before its application, with the tuning process being dependent on the material, interatomic potential, crystal structure, and thermodynamic conditions (i.e., simulation temperature). The PTM accuracies shown on Table \ref{tab:accuracies} correspond to the performance of PTM using a set of highly optimized parameters tuned to work individually with each material and temperature in the benchmark data set (see Appendix \ref{app:PTM_optimization} for the PTM optimization details). Another point worth noticing is that PTM cannot be globally optimized to work with multiple crystalline phases or materials. For example, the optimal parameters for Ge (cd) result in an accuracy of $99.2\%$ for Ge but only an accuracy of $8.3\%$ for NaCl (sc), but PTM can achieve up to $94.6\%$ accuracy if optimized specifically for NaCl. This can be a complication when working with multiple crystalline phases such as, for example, when investigating the interface between dissimilar crystal structures. The optimal PTM parameters are also not transferable across different materials with the same crystal structure. For example, the optimal parameters for Ge (cd) decreases the accuracy of Si (also cd) from $99.9\%$ to only $55.9\%$. DC3 is simultaneously optimized for each crystal structure and does not suffer from this problem, i.e., DC3 can be applied with its optimal accuracy to any number of crystal structures simultaneously.

  \subsection{Temperature dependence} The accuracies shown on Table \ref{tab:accuracies} correspond to MD simulations performed at a single temperature, namely the melting point of each respective material. Yet, it is desirable for crystal-structure classification methods to perform well under the presence of a variety of thermal noise amplitudes (i.e., temperatures). Further MD simulations were performed for all materials in the benchmark data set  in order to evaluate the DC3 model transferability with respect to temperature. These simulations covered the entire temperature range from $0.04\,T_\text{m}$ to $T_\text{ml}$, where $T_\text{ml}$ is the highest temperature at which the crystal phase could be equilibrated (see Appendix \ref{app:MD} for details on the calculation of $T_\text{ml}$), shown on Table \ref{tab:materials_and_potentials}. The temperature dependence of the accuracy of DC3 and the heuristic methods is shown in Fig.~\ref{fig:temperature_dependence} (see Fig.~S1 \footnote{See Supplemental Material at [URL] for supplementary figures.} for the results of the other materials on Table \ref{tab:materials_and_potentials}). It can be seen in this figure that DC3 remains the best performing method for the entire range of temperatures. Notice that for H$_2$O (hd) the trends on Table \ref{tab:accuracies} seem to change at $T \approx T_\text{ml}$, where the accuracy of DC3 seemly becomes superior to PTM due to better robustness against thermal noise (i.e. slower decrease in accuracy with temperature). It can be seen in Fig.~\ref{fig:temperature_dependence} that all methods present a general trend: lower accuracy as temperature increases. Such a trend is due to the fact that structural distortions caused by thermal noise make it more difficult to discern between the crystal structures. This can be seen in Fig.~\ref{fig:tsne}c, where the temperature-resolved tSNE analysis of the MD simulations of Ti (hcp) shows that data points corresponding to high temperatures are the ones farthest from the perfect crystalline structure, and therefore closer to areas of the feature space occupied by other crystal structures.
  \begin{figure*}[tb]
      \centering
      \includegraphics[width=\textwidth]{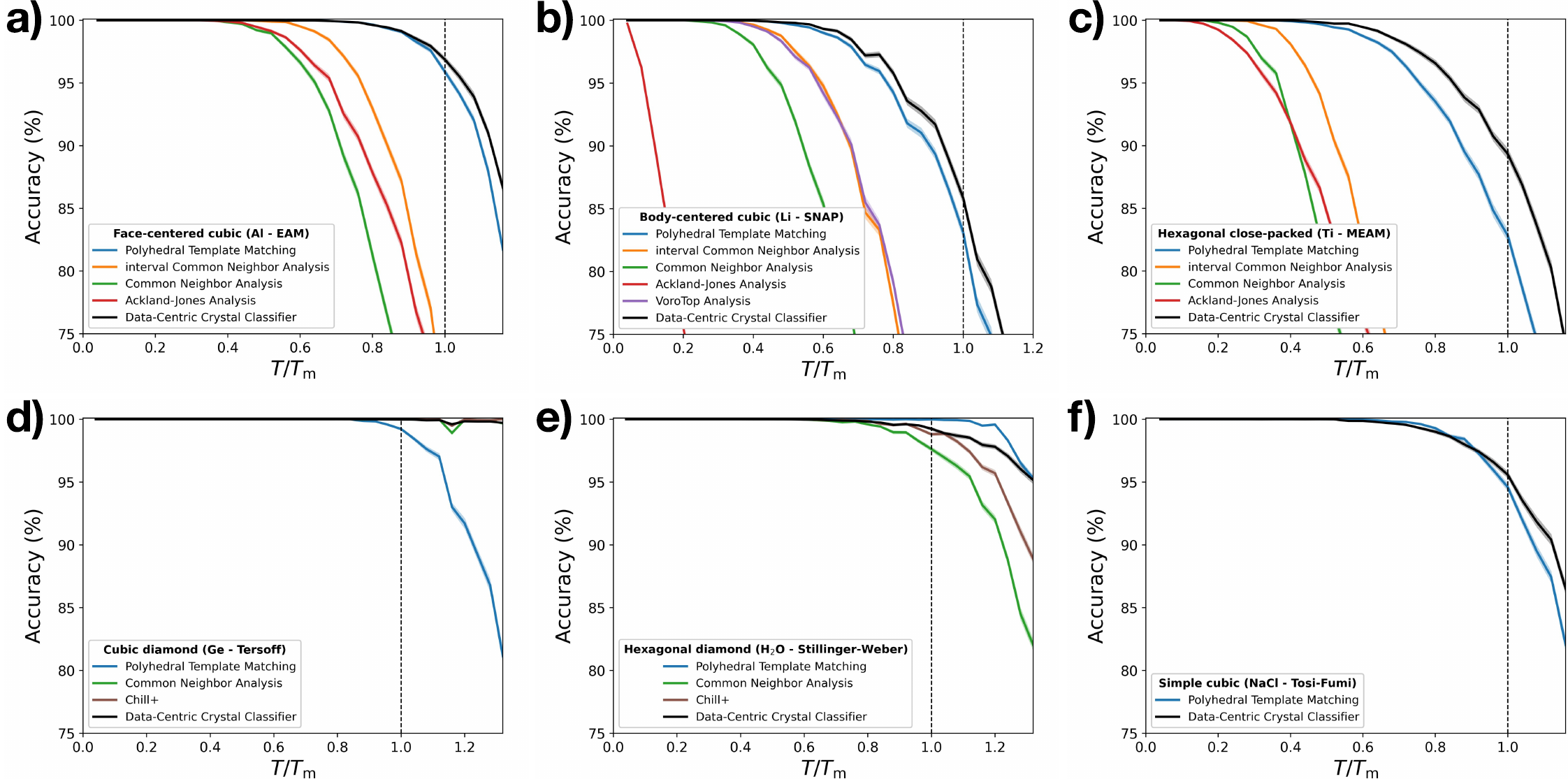}
      \caption{Accuracy as a function of the homologous temperature as evaluated on materials of the benchmark data set (Table \ref{tab:materials_and_potentials}). Lines represent the mean accuracy, while the shadows show the $95\%$ confidence interval around the mean (see Appendix \ref{app:error} for details on the confidence interval calculation). The temperature of the MD simulations ranged from $0.04\,T_\text{m}$ to $T_\text{ml}$, where $T_\text{ml}$ is the highest temperature at which the crystal structure could be equilibrated. The vertical dashed line marks the melting temperature. The general trend of lower accuracy as temperature increases is due to structural distortions caused by thermal noise. Notice that not all heuristic methods are capable of identifying all crystal structures. Similar results for other materials on Table \ref{tab:materials_and_potentials} are shown in Fig.~S1 \cite{Note1}.}
      \label{fig:temperature_dependence}
  \end{figure*}
  
  The tSNE analysis of the MD-generated data set is shown in Fig.~\ref{fig:tsne}b and it can be compared to the tSNE distribution of the synthetic data set in Fig.~\ref{fig:tsne}a. The similarities between the two distributions is another evidence that our procedure to build a synthetic data set was successful in creating an effective training set for the ML classifier. 
  
  \subsection{Generalization to chemically complex structures \label{sec:chemically_complex}} 
\begin{figure}[tb]
    \centering
    \includegraphics[width=0.45\textwidth]{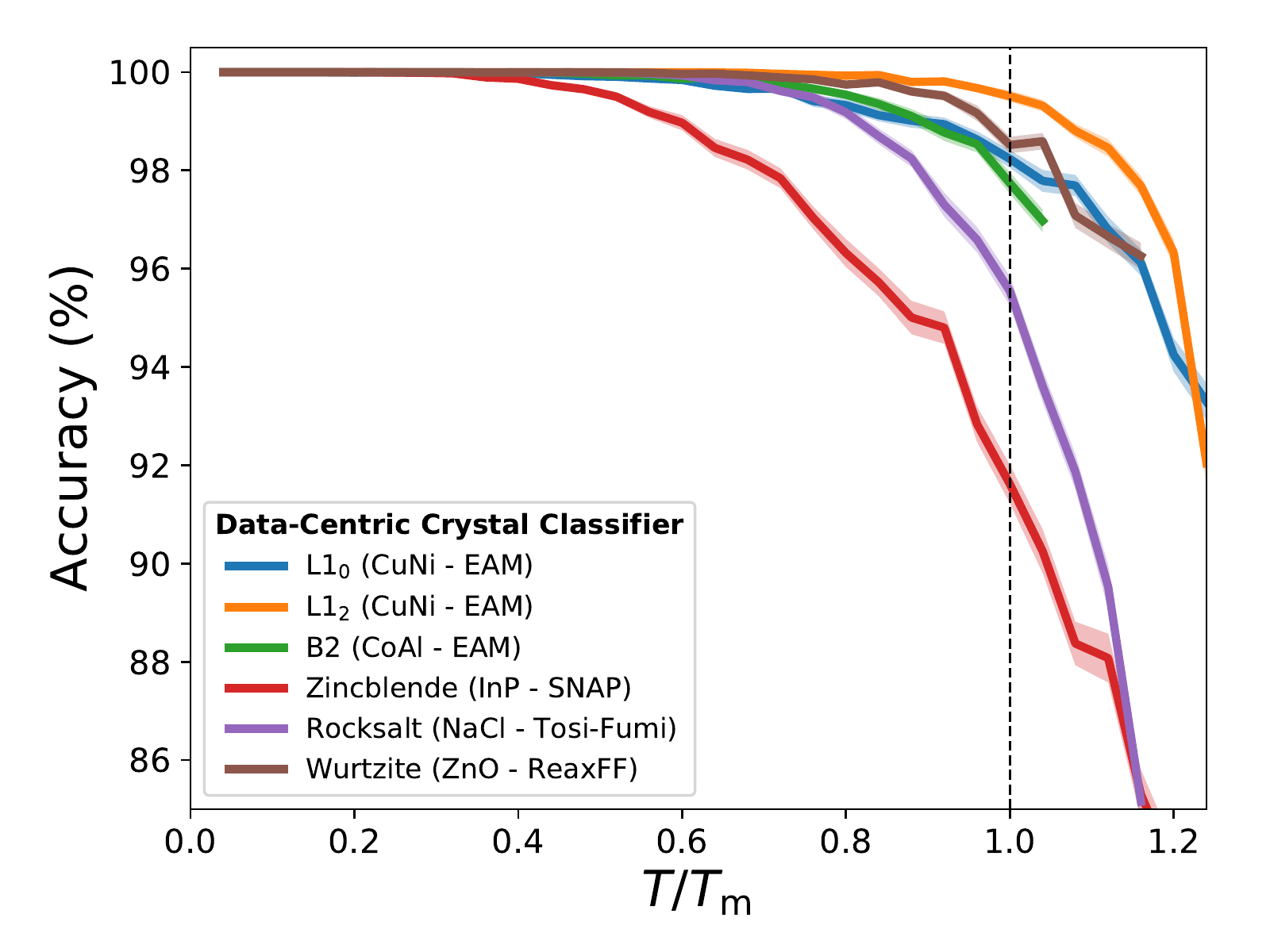}
    \caption{\label{fig:chemical_complex} Accuracy of the DC3 method for binary compounds as a function of the homologous temperature. Lines represent the mean accuracy, while the shadows show the $95\%$ confidence interval around the mean (see Appendix \ref{app:error} for details on the confidence interval calculation). The DC3 approach was capable of retaining a similar level of accuracy for chemically complex structures as those observed for single-element structures. Extension of DC3 to multi-element systems was accomplished by simply augmenting the feature vector $\v{x}_i$ to account for different chemical species.}
  \end{figure}

  The DC3 model developed so far is only capable of classifying crystal structures of materials composed of a single element (benchmark data set on Table \ref{tab:materials_and_potentials}). This limitation can be overcome by appropriately modifying the feature vector $\v{x}_i$. In this section we demonstrate how this can be accomplished for binary compounds: A$_n$B$_m$. The approach chosen here is to augment the feature vector definition such that $\v{x}_i = \left( \v{x}^\text{A}_i, \v{x}^\text{B}_i \right)$, where $\v{x}^\text{X}_i$ is the same feature vector defined for single-element materials but computed using only those neighbors of atom $i$ that belong to the chemical species $X$. In order to test this approach we have performed MD simulations of six binary compounds (listed on Table \ref{tab:materials_and_potentials}) at temperatures ranging from $T_\text{m}$ to $T_\text{ml}$. The accuracy as a function of temperature is shown in Fig.~\ref{fig:chemical_complex}, where it can be seen that the DC3 approach retains similar levels of accuracy as those observed for single-element structures. The average accuracy for the corresponding liquid phases was $99.4\% \pm 0.1\%$. See Appendix \ref{app:binary} for more details on the DC3 generalized algorithm.
 
  The fact that DC3 performance is similar for single-element materials and binary compounds warrants some observations since the task of creating heuristic rules to discern binary-compound structures is more difficult than for single-element materials due to the increased complexity added by the presence of multiple elements. Hence, it is unexpected that this does not seem to be the case for an ML-based data-centric approach. For example, the L1$_0$ and L1$_2$ crystal structures are both alloys obtained from the fcc structure by simple chemical substitution performed at certain crystal lattice points. Yet, DC3 is capable of discerning between these two structures with perfect accuracy within the standard error (i.e. $\le 0.1\%$), as can be seen in the confusion matrix in Fig.~S6 \cite{Note1}. In fact, all four chemically complex structures with cubic symmetry show less than $0.4\%$ false-positive errors. These excellent accuracies suggest that DC3 is capable of easily assimilating the extra information contained in the description of chemically complex structures. The same chemical complexity that makes the creation of heuristic rules more arduous for human intuition.

\section{Discussion and Conclusions}
\begin{figure*}[tb]
    \centering
    \includegraphics[width=\textwidth]{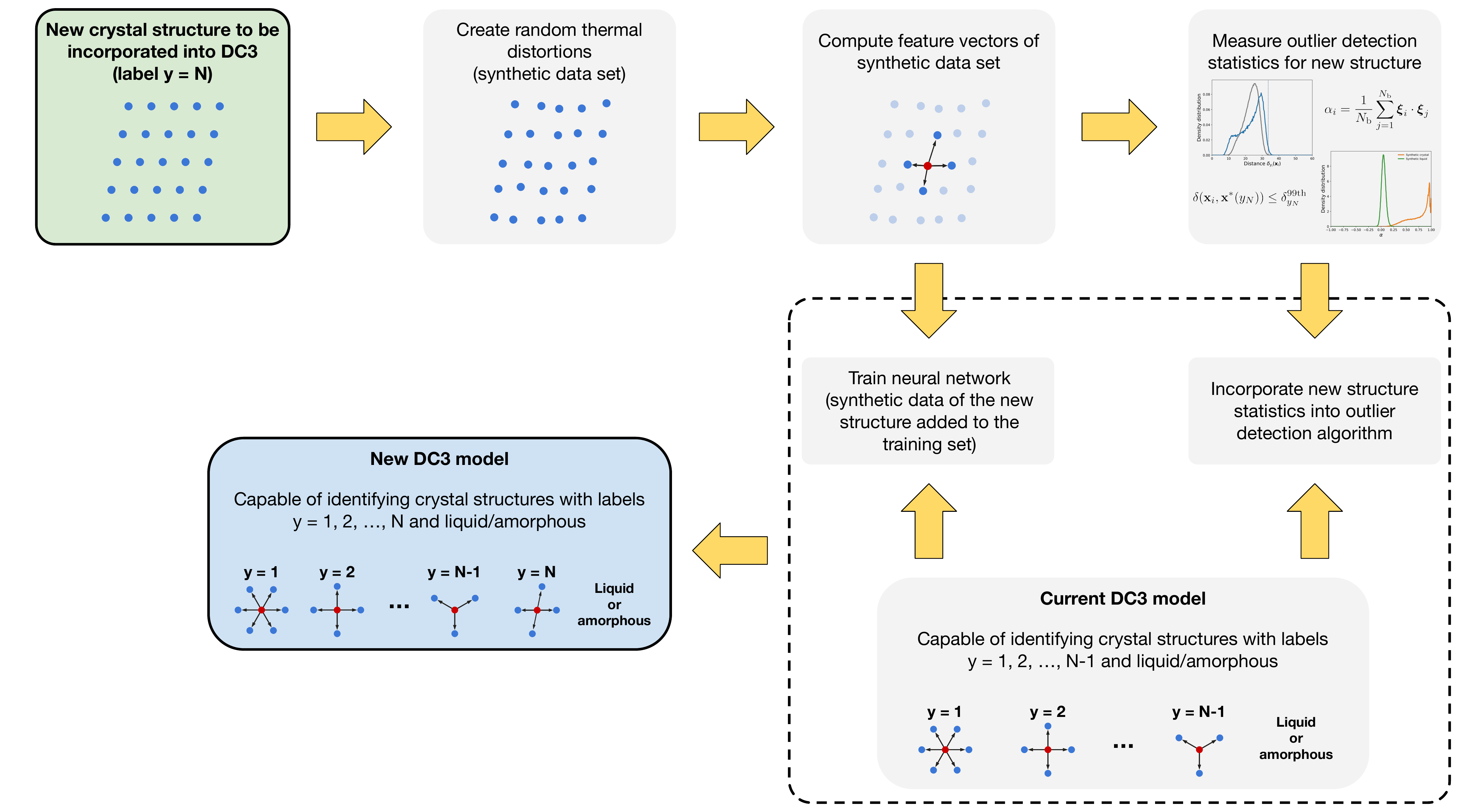}
    \caption{\label{fig:new_structure} Flowchart illustrating the steps required to incorporate a new crystal structure with label $y=N$ into an already existing DC3 model capable of identifying  crystal structures with labels $y=1,2,\ldots,N-1$. The DC3 framework can be systematically generalized to novel crystal structures without the need to perform any materials simulation, making it especially suitable to high-throughput approaches. Notice how the pipeline does not require predefined rules to be derived for each new structure.}
  \end{figure*}

  One of the distinguishing properties of the DC3 framework is that it can be systematically generalized to novel crystal structures, as illustrated in Fig.~\ref{fig:new_structure}. This follows from its streamlined pipeline (Fig.~\ref{fig:framework}) that does not require predefined rules to be derived for each new structure. Instead, a synthetic data set is automatically generated from the description of the new crystal structure and a neural network learns structural distortions patterns from the data. With this data-centric approach DC3 can be easily generalized to an arbitrary number of complex crystal structures. It has been shown here that new DC3 models with extended capabilities can be generated by simply extending the feature vector $\v{x}_i$, such as adding species-sensitive features in order to classify chemically complex structures. Extending the feature vector is a much simpler task than creating new heuristic rules to achieve the same capabilities. Thus, DC3 is a natural algorithm to employ when flexibility to adapt to new material structures is imperative, such as in the high-throughput frameworks \cite{aflow,high_throughput,mp,fireworks} pervasive in materials science nowadays.

  The approach developed here is also the only crystal-structure classification algorithm capable of discerning unknown crystal structures from amorphous motifs (i.e., noncrystalline phases). All heuristic methods bundle these two categories together and only differentiate them from known crystal structures. This capability makes DC3 useful in simulations of crystal growth and other materials synthesis processes where the formation of novel crystal structures might go unnoticed if such structure is not already contained in the method's database. With DC3 this problem would not arise because a novel crystal phase would be identified as such, and not as an amorphous structure that might be easily mistaken as not part of the material being synthesized. For example, during crystal growth from the melt the nucleating crystallite is embedded in a matrix of its own liquid phase. In this case DC3 easily identifies the new crystallite (as a known or unknown crystal structure) while recognizing the liquid phase as a noncrystalline phase.

  Identification of crystal defects is an important problem in computational materials science. Methods to detect defects invariably begin \cite{dxa,tim_rupert,grain_rotation} by first identifying the atoms belonging to perfect crystal structure, since those need not to be included in the search and classification of crystal defects. This requirement has limited the applicability of crystal defect identification methods to lattices for which crystal classification algorithms exist. This limitation can be lifted with DC3 because of its ability to discern unknown crystal structures from amorphous motifs. With DC3 one will be able to identify crystal defects in any crystal structure, even those unknown to DC3.

  More generally, the approach developed here is applicable not only to crystal structures, but also to any general atomic motif that needs to be identified in the presence of thermal noise. Such as for example, local chemical order in high-entropy alloys \cite{hea}, unusual coordination in supercooled liquids \cite{water_valeria}, solvation shells in various aqueous solutions \cite{dc_solvation}, grain boundary structures \cite{gb_1,gb_2}, and structural features at surfaces \cite{steps_1,steps_2}. The only requirement is the creation of an appropriate synthetic data set for training DC3 framework to identify the atomic motif of interest.

  In conclusion, we have created a data-centric framework for crystal-structure classification in atomistic simulations. Our approach does not rely on expert knowledge to derive rules for discerning crystal structures, instead the task of deriving heuristic rules is delegated to an ML algorithm that learns the optimal patterns of classification from an unbiased and material-independent synthetic data set. The entire process of generalization to new structures can be easily automated to run without human intervention, a capability that sets this approach apart from heuristic structure classification algorithms. We have also created the first statistically rigorous benchmark data set to compare performance of different methods. Using this benchmark data set it was shown that DC3 has better performance than any of the most popular heuristic algorithms. The data-centric framework developed here has proven to be a flexible template on which more specialized methods can easily build upon for niche applications. Because of its data-centric format, extension of DC3 capabilities can be performed using a variety of state-of-the-art ML and data-science methods.
  
\begin{acknowledgments}
  This work was supported by the Department of Energy National Nuclear Security Administration under Award Number DE--NA0002007, National Science Foundation grants DMREF--1922312 and CAREER--1455050, and the Air Force Office of Scientific Research under award number FA9550-20-1-0397.
\end{acknowledgments}

\section*{Competing interests}
  The authors declare no competing interests.

\section*{Code and data availability}
  The code and scripts used to generate the results in this paper can be downloaded from the following repository: \url{https://github.com/freitas-rodrigo/DC3}. In addition, any data that supports the findings of this study are available from the corresponding author upon reasonable request.

\clearpage
\appendix

\section{Molecular Dynamics simulations \label{app:MD}}
  The MD simulations were performed using the Large-scale Atomic/Molecular Massively Parallel Simulator (LAMMPS \cite{LAMMPS}). Atomic forces and energies were described using the following interatomic potentials:  Finnis-Sinclair embedded-atom model (EAM-FS \cite{eam_1,eam_2,eam_fs}) for Fe \cite{eam_Fe}, Al \cite{eam_Al}, Mg \cite{eam_Mg}, CuNi \cite{eam_CuNi}, and CoAl \cite{eam_CoAl}, modified embedded-atom model (MEAM \cite{meam_1,meam_2}) for Ti \cite{meam_Ti}, Stillinger-Weber \cite{sw} for a coarse-grained monoatomic model of H$_2$O \cite{sw_mW}, environment-dependent interatomic potential (EDIP \cite{edip}) for Si, Tosi-Fumi model \cite{tosi_fumi_1,tosi_fumi_2} for NaCl, spectral neighbor analysis potential (SNAP \cite{snap}) for Li \cite{snap_Li} and InP \cite{snap_InP},  Lennard-Jones \cite{lj_1,lj_2} for Ar \cite{lj_Ar}, Tersoff \cite{tersoff} for Ge \cite{tersoff_Ge}, and ReaxFF \cite{reaxff_1,reaxff_2} for ZnO \cite{reaxff_ZnO}.

  The timestep for all simulations was $1\,\text{fs}$. The Bussi-Donadio-Parrinello thermostat \cite{bdp} was employed with a relaxation time of $0.1\,\text{ps}$. In order to maintain the system at zero pressure we employed an isotropic Nosé-Hoover chain barostat \cite{nh_1,nh_2,nh_3} with chain length of three and relaxation time of $1\,\text{ps}$. The system size was chosen such that it contained at least $17000$ atoms while maintaining the system dimensions as close to a cube as possible.

  For each crystal structure (and corresponding interatomic potential) the system was initialized with atoms in a perfect crystal structure with lattice parameter corresponding to the zero temperature equilibrium value. The system was equilibrated at the target temperature and zero pressure for $10\,\text{ps}$ followed by a period of $90\,\text{ps}$ during which snapshots of the atomic coordinates were collected every $10\,\text{ps}$.

  Simulations were run for each structure at temperatures ranging from $0.04\,T_\text{m}$ to $1.60\,T_\text{m}$ in intervals of $0.04\,T_\text{m}$. In order to determine if the system had melted over the course of a simulation the atomic coordinates of the last snapshot of each simulation were relaxed to their local energy minima using the conjugate-gradient \cite{cg} algorithm with a quadratic line search and maximum atom displacement of $0.1\Ang$. We considered the relaxation to have converged once the line search backtracks to zero distance (i.e., the energy and force tolerances are zero and there is no limit on the number of iterations).

  The final data set obtained from MD simulations contained $17000$ data points per crystal structure per temperature. The accuracy statistics of different classification methods was computed employing only data from temperatures for which the system was $100\%$ crystalline at the end of the MD simulation. Notice that crystalline phases can be metastable and therefore retain their crystal structure above $T_\text{m}$. Nevertheless, each material has a limit of metastability marked by a temperature $T_\text{ml} \ge T_\text{m}$ above which the crystal phase cannot be equilibrated anymore. Here we have estimated $T_\text{ml}$ for each material as the highest temperature at which the last snapshot of the MD simulation relaxed back to its perfect crystal structure. Data points at $T > T_\text{m}$ are invaluable when benchmarking the accuracy of crystal-structure classification methods because they are the most challenging to classify due to the large thermal fluctuations present.

  The statistics of the liquid phase was computed for all materials using data from $T = 1.60\,T_\text{m}$, a temperature at which all materials had completely melted after the initial equilibration time.
  
  \section{Synthetic data creation \label{app:synthetic}}
  Synthetic data for each crystal structure was constructed by initializing a system with atoms at their perfect crystal structure positions and applying a random displacement to each atom. These random displacements are performed uniformly over a sphere of radius $r$ by selecting the spherical coordinates of the displacement vector as follows. The azimuthal angle $\phi$ is chosen randomly over a uniform distribution with range $[0,2\pi)$, while the polar angle $\theta$ is chosen such that $\cos\theta$ is uniformly distributed over $[-1,1)$. The radial distance $r$ is chosen such that $r^3$ is uniformly distributed, while $r$ falls inside the interval $[0,\alpha d)$, where $d$ is the distance to the first neighbor in the perfect crystal structure and $\alpha$ is a parameter used to control the magnitude of the displacements. In order to mimic structures at different temperatures we employed $40$ different values of $\alpha$ uniformly distributed in the range $[0.01,0.25]$. The final synthetic data set contained $69000$ unique data points per crystal structure.

  \section{Feature vector construction \label{app:feature_vector}}
  Each atom had its local neighborhood described by a set of features composed of the two families of functions shown in Eqs.~\eqref{eq:steinhardt} and \eqref{eq:rsf}. It is clear from these equations that Steinhardt parameters capture angular information about the local structure, while RSF account for the radial information. Each RSF evaluates the density of atoms at a distance $r$ away from the central atom. If a neighbor's radial distance is different from $r$ its contribution to the density is controlled by $\sigma$. Meanwhile, for each value of $\ell$ the Steinhardt parameter is sensitive to a different orientational symmetry present in the local neighborhood of atom $i$ as defined by its first $N_\text{b}$ neighbors.  

  Next we describe how the free parameters in Eqs.~\eqref{eq:steinhardt} and \eqref{eq:rsf} were selected. For $Q_\ell^{N_\text{b}}(i)$ the free parameters  are $\ell$ and $N_\text{b}$. Traditionally, $\ell = 4$ and $6$ are often employed with mild success to differentiate certain crystal structures, mostly because they allow the classification to be done by visual inspection of a two-dimensional plot. Here we take the data-centric route and employ $\ell = 1, 2, 3, 4, \ldots, 15$ ($\ell = 0$ results in constant spherical harmonics), leaving the task of extracting information from the high-dimensional data created to the ML algorithm. The $N_\text{b}$ parameter was picked in a similar spirit. The number of nearest neighbors in each crystal structure considered here are $6$ for sc, $8$ for bcc, $12$ for fcc and hcp, $16$ for cd and hd. Hence, we chose $N_\text{b} = 2, 3, 4, \ldots 16$ ($N_\text{b} = 1$ does not contain any local-symmetry information) for each value of $\ell$. Notice that the cd and hd structures have $4$ identical nearest neighbors, thus it is necessary to employ the $12$ second nearest neighbors in order to differentiate these two structures.
  
  The free parameters for $G_{r,\sigma}^{N_\text{b}}(i)$ ($r$ and $\sigma$) are computed as follows. Notice that in order to render $\v{x}_i$ independent of the material lattice constant the parameters $r$ and $\sigma$ are not preselected before the local-structure characterization takes place. Instead, these parameters are determined with regard to a local-distance metric calculated only during the evaluation of $\v{x}_i$, which effectively renders $\v{x}_i$ independent of the magnitude of the lattice constant (this approach was inspired by the adaptive cutoff method developed by \citet{stuko_adaptive}). The local-distance metric for a given atom $i$ and a given number of nearest neighbors $N_\text{b}$ is chosen as the the average radial distance of the first $N_\text{b}$ nearest neighbors of $i$:
  \[
    \left<r_{ij}\right>_{N_\text{b}} = \frac{1}{N_\text{b}} \sum_{j=1}^{N_\text{b}} r_{ij}.
  \]
  Hence, the values of $r$ employed for atom $i$ are $r / \left<r_{ij}^{N_\text{b}} \right>= 0.85, 0.90, 0.95, 1.00, 1.05, 1.10, 1.15$, while $\sigma(i) = 0.05 \left<r_{ij}^{N_\text{b}}\right>$. Notice that the summation in Eq.~\eqref{eq:rsf} is over all atoms within a cutoff radius $r_\text{cut}$, instead of a summation over the first $N_\text{b}$ neighbors. The parameter $N_\text{b}$ is only employed to determine $r$ and $\sigma$ for each atom. The value of $r_\text{cut}$ is chosen such that contributions of neighbors beyond the cutoff are negligible: $r_\text{cut} = \max [r(i)] + 4 * \max [ \sigma(i)]$, where the $\max$ function is taken over all atoms in the system for a given $N_\text{b}$.

  Finally, the feature vector $\v{x}_i$ for atom $i$ is constructed as follows:
  \begin{equation}
    \label{eq:x}
    \begin{matrix} 
       \v{x}_i = \Big[\!\!\!\! &            Q_1^2, &            Q_2^2, &            Q_3^2, & \ldots \, , &         Q_{15}^2, & \\[6pt] 
                               &            Q_1^3, &            Q_2^3, &            Q_3^3, & \ldots \, , &         Q_{15}^3, & \\[6pt]
                               &            Q_1^4, &            Q_2^4, &            Q_3^4, & \ldots \, , &         Q_{15}^4, & \\[6pt]
                               &       \vdots \, , &       \vdots \, , &       \vdots \, , & \vdots \, , &       \vdots \, , & \\[6pt]
                               &         Q_1^{16}, &         Q_2^{16}, &         Q_3^{16}, & \ldots \, , &      Q_{15}^{16}, & \\[6pt]
                               &    G_{r,0.85r}^2, &    G_{r,0.90r}^2, &    G_{r,0.95r}^2, & \ldots \, , &    G_{r,1.15r}^2, & \\[6pt]
                               &    G_{r,0.85r}^3, &    G_{r,0.90r}^3, &    G_{r,0.95r}^3, & \ldots \, , &    G_{r,1.15r}^3, & \\[6pt]
                               &       \vdots \, , &       \vdots \, , &       \vdots \, , & \vdots \, , &       \vdots \, , & \\[6pt]
                               & G_{r,0.85r}^{16}, & G_{r,0.90r}^{16}, & G_{r,0.95r}^{16}, & \ldots \, , & G_{r,1.15r}^{16}  & \!\!\!\! \Big],
    \end{matrix}
  \end{equation}
  where the $(i)$ dependence inside the vector was omitted for clarity.
  
  Note that, in total, our feature vector has length 330. Given this magnitude of features, significant training data (on the order of at least 50000 examples) must be used in order to avoid overfitting. Since, generating this amount of training data using MD would be inefficient and even infeasible for more chemically complex systems, our synthetic method may be the only way to train this kind of framework.

  Finally, a linear transformation was applied to the collection of feature vectors in the synthetic data set such that each component of $\v{x}_i$ had zero mean and standard deviation of one. The same linear transformation was subsequently applied to the MD data sets (benchmark data set on Table \ref{tab:materials_and_potentials}).

  \section{Machine Learning model training \label{app:training}}
  The synthetic data set (consisting of $414000$ data points) was employed to parametrized a multi-class feed-forward neural network composed of three hidden layers with $100$ rectified linear units and softmax output one \cite{deep_learning_book,sklearn}. Training was performed using the early stopping strategy \cite{deep_learning_book}: $10\%$ of the training set was set aside and used as validation, training stopped when the validation score did not improve by at least $10^{-4}$ for $10$ epochs. Optimization of the log-loss function was carried out using the Adam \cite{adam} algorithm with $\beta_1 = 0.9$, $\beta_2 = 0.999$, $\epsilon = 10^{-8}$, learning rate initialization of $5\times10^{-3}$, minibatches of size $200$, and a L2-regularization term with parameter $10^{-4}$. The model parametrization strategy described above was decided based on a hyperparameter optimization procedure shown in Fig.~S2 \cite{Note1}.

  \section{Outlier detection algorithm \label{app:outlier}}
  The outlier detection algorithm has two different objectives. First it must determine whether a feature vector $\v{x}_i$ corresponds to a crystal structure or an amorphous atomic motif. If this first assessment concludes that $\v{x}_i$ corresponds to a crystal structure, then the second goal of the outlier detection algorithm is to determine whether the crystal structure is known (i.e., is included in the ML classifier database) or unknown.
  
  In order to determine whether $\v{x}_i$ corresponds to a crystal structure or an amorphous motif the information contained in $\v{x}_i$ is used to construct the following vector:
  \[
    \tilde{\gv{\xi}}_i = \Big[ \v{q}_4^{N_\text{b}}(i), \v{q}_6^{N_\text{b}}(i), \v{q}_8^{N_\text{b}}(i), \v{q}_{12}^{N_\text{b}}(i) \Big]
  \]
  where $\v{q}_\ell^{N_\text{b}}(i)$ is defined as:
  \[
    \v{q}_\ell^{N_\text{b}}(i) = \Big[q_{\ell, m=-\ell}^{N_\text{b}}(i), q_{\ell, m=-\ell+1}^{N_\text{b}}(i), \, \dots  \, , q_{\ell, m=\ell}^{N_\text{b}}(i) \Big],
  \]
  with $q_{\ell, m}^{N_\text{b}}(i)$ given in Eq.~\eqref{eq:q_lm}. Finally, vector $\gv{\xi}_i$ is the normalized version of $\tilde{\gv{\xi}}_i$: 
  \begin{equation}
    \label{eq:xi}
    \gv{\xi}_i = \frac{\tilde{\gv{\xi}}_i}{\left| \tilde{\gv{\xi}}_i \right|}.
  \end{equation}
  Vector $\gv{\xi}_i$ is such that the more similar the structure surrounding atoms $i$ and $j$ is, the more parallel vectors $\gv{\xi}_i$ and $\gv{\xi}_j$ will be. For example, in a perfectly undisturbed Bravais lattice $\gv{\xi}_i \cdot \gv{\xi}_j = 1$ for any two pairs of atoms $i$ and $j$, while for a system composed of a truly random distribution of atoms $\left< \gv{\xi}_i \cdot \gv{\xi}_j\right> = 0$, where the average $\left< \ldots \right>$ is over all atoms in the system. Finally, the structure coherence parameter is defined as
  \[
    \alpha_i = \frac{1}{N_\text{b}} \sum_{j=1}^{N_\text{b}} \gv{\xi}_i \cdot \gv{\xi}_j.
  \]
  
  Using the synthetic training set the distribution of $\alpha_i$ was evaluated for all crystal structures. Similarly, another data set composed of randomly distributed atoms, meant to mimic amorphous and liquid materials, was employed to determine the $\alpha_i$ distribution for amorphous motifs. Based on these two distributions (shown in Fig.~S3 \cite{Note1}) the optimal cutoff value $\alpha_\text{cutoff} = 0.196$ was determined such as to maximize the probability of correctly classifying an atom as belonging to a crystal structure or to an amorphous motif: if $\alpha_i \ge \alpha_\text{cutoff}$ then atom $i$ is determined to be crystalline, otherwise it is surrounded by an amorphous arrangement of neighbors.
  
  The above definition of the structure coherence factor $\alpha_i$ warrants some observations. First, notice that our definition of $\alpha_i$ and $\gv{\xi}_i$ is a direct generalization of concepts introduced by the work of  \citet{alpha_frenkel}. We have first attempted to define $\gv{\xi}_i = \v{x}_i$, but this definition led to poor accuracy of the outlier detection algorithm. One of the most important properties of $\v{x}_i$ is that it is rotationally invariant, which is only the case because the definition of $Q_\ell^{N_\text{b}}(i)$ involves rotationally invariant combinations of $q_{\ell,m}^{N_\text{b}}(i)$. This is a desirable property for the ML classifier because a feature vector corresponding to a specific crystal structure should be the same no matter the (physically arbitrary) orientation of the system in space. But this is not a requirement when trying to discern if the local structure surrounding atom $i$ is crystalline or not, because $\gv{\xi}_i$ is only compared to a few nearest neighbors. Thus, in defining $\gv{\xi}_i$ we forgo the rotationally invariant requirement in order to retain the angular information that is lost when the modulus of $q_{\ell,m}^{N_\text{b}}(i)$ is taken. We have also observed that the most important contributions to $\alpha_i$ came from $\ell = 4, 6, 8,$ and $12$, since no statistically measurable improvement was observed by employing more values of $\ell$ or the RSF (Eq.~\eqref{eq:rsf}). Hence, $\gv{\xi}_i$ took the more compact definition presented in Eq.~\eqref{eq:xi}, which does not require any additional computation since all information needed to evaluate $\gv{\xi}_i$ was already computed when evaluating $\v{x}_i$. Finally, the $N_\text{b} = 16$ value was employed since this is the maximum number of first neighbors of all structures considered on Table \ref{tab:materials_and_potentials}.
  
  The outlier detection algorithm ends if the analysis of $\alpha_i$ determined that the local structure surrounding atom $i$ is amorphous. But, if the analysis above determined the local structure to be crystalline the algorithm must now decide whether this is a known crystal structure or not. In this case the next step is to apply the ML classifier to predict a label $\tilde{y}_i$ to this feature vector. Then it must be statistically decided whether label $\tilde{y}_i$ is appropriate or $\v{x}_i$ corresponds instead to an unknown crystal structure. In order to make that decision notice that each crystal structure has a feature vector that is special compared to all others: the one corresponding to the undistorted crystal structure (i.e., no thermal noise), shown in Fig.~\ref{fig:tsne}. For a crystal structure with label $\tilde{y}_i$ this special feature vector $\v{x}^*(\tilde{y}_i)$ is the perfect example of the crystal structure. Hence, the distance $\delta(\v{x}_i,\v{x}^*(\tilde{y}_i))$ from $\v{x}_i$ to $\v{x}^*(\tilde{y}_i)$ is evaluated and compared to the location of the $99$th percentile ($\delta_{\tilde{y}_i}^{99\text{th}}$) of the distance density distribution as obtained from the synthetic data set (see Fig.~\ref{fig:distance_density} for an example for the case of the hcp structure). If $\delta(\v{x}_i,\v{x}^*(\tilde{y}_i)) \le \delta_{\tilde{y}_i}^{99\text{th}}$ the label $\tilde{y}_i$ is considered appropriate, otherwise $\v{x}_i$ is determined to belong to an unknown crystal structure.

  \section{Benchmark of crystal classification methods \label{app:benchmark}}
  In order to assess the quality of the DC3 method we have also computed the accuracy of different algorithms of crystal classification available in the literature. The methods considered here were: Polyhedral Template Matching (PTM \cite{ptm}), Common-Neighbor Analysis (CNA \cite{cna_1,cna_2}), interval Common-Neighbor Analysis (iCNA \cite{icna}), Ackland-Jones Analysis (AJA \cite{aja}), VoroTop \cite{vorotop}, and Chill+ \cite{chillp}.

  Each of these methods require a different set of parameters in order to perform optimally. The PTM optimization process is intricate and we devote the next section to describe it. The CNA and iCNA methods were employed here in conjunction with the adaptive approach developed in Ref.~\onlinecite{stuko_adaptive}, which has the dual benefits of having no global adjustable parameter and also performing better than the standard fixed-cutoff variant. The AJA method does not require any adjustable parameter. The VoroTop method requires the application of a filter to specify the structure types. Here we have chosen the most general filter available: FCC-BCC-ICOS-HCP. The Chill+ method requires a cutoff radius for the first-neighbor bond, which we have set to $3.4\Ang$ (fcc), $3.0\Ang$ (bcc), $3.5\Ang$ (hcp), $2.9\Ang$ (cd), $3.5\Ang$ (hd), $3.5\Ang$ (sc). Notice that Chill+ can only classify cd and hd crystal structures, but in order to obtain good accuracy for liquid classification of the other structures we also optimized the cutoff for their respective materials.

  Not all structures can be classified using the methods considered here, with the single exception of PTM, which works for all six crystal structures. The CNA and iCNA methods can classify fcc, bcc, hcp, cd, and hd. The AJA and VoroTop methods work for fcc, bcc, and hcp. The Chill+ works for cd and hd.

  In order for the accuracy comparison among the methods and against DC3 to be considered fair we took the following steps. The option to classify icosahedral structures was turned off for the CNA, iCNA, and AJA (the VoroTop filter does not allow for ignoring icosahedral structures but we did verify that the amount of icosahedral misclassification is not statistically relevant). The option to identify first and second neighbors in cd and hd structures was turned off for the CNA and iCNA.

  The statistics of each method were collected by applying them to the crystal and liquid data sets obtained from MD simulations. The synthetic data set was used solely in the parametrization and model selection for the DC3 method.

  \section{Optimization of the PTM method \label{app:PTM_optimization}}
  During the process of classification the PTM method estimates how close a certain atom neighborhood is to a perfect crystal structure (i.e., the similarity between both sets of points) by evaluating the root-mean-square deviation (RMSD) between the data point and the perfect structure template. The classification decision is made by computing the RMSD between the data point and all structures considered, and then choosing the structure that results in the lowest RMSD. A data point is determined to not belong to any of the structures considered if the all RMSDs computed are above a predetermined cutoff value. This cutoff value is temperature and material dependent and must be chosen carefully prior to the application of the PTM in order for the method to perform optimally. For example, consider a system that is half crystalline and half liquid. If the RMSD cutoff is too large many liquid atoms can be mistakenly classified as crystalline, while a too small value will result in many atoms belonging to the crystal classified as liquid.

  Using the data set obtained from MD simulations we have computed the density distribution of the RMSDs obtained for each material at $T = T_\text{m}$ (see Fig.~S7 \cite{Note1}). It is clear from Fig.~S7 \cite{Note1} that each material has a different optimal RMSD cutoff value. For example, Si (cd) has an optimal cutoff of $0.1425$ while NaCl (sc) requires a cutoff of at least $0.2145$. Hence, differently from DC3 the PTM method cannot be simultaneously optimized for all materials. Moreover, in Fig.~S8 \cite{Note1} we show that the accuracy of classification of the liquid phase is also strongly dependent on the choice of the RMSD cutoff.

  In order to obtain the optimal classification accuracy for all materials in their crystalline and liquid phases we optimized the PTM method individually for each material as follows. First density distribution of RMSD values for each material's crystal (at $T = T_\text{m}$) and liquid phases (at $T = 1.6T_\text{m}$) was evaluated (see Fig.~S7 \cite{Note1}). Then the accuracy of the PTM method was evaluated for RMSD values in the range from $0.0$ to $0.6$ in intervals of $0.003$ both phases (crystal and liquid). For each material the average between the accuracy of the crystal and liquid was computed (Fig.~S8 \cite{Note1}) and the RMSD at the maximum average accuracy was employed as RMSD cutoff.
   
  This trade-off between the accuracy of the crystal structures and the accuracy of the liquid (or unknown structures) can be seen from a data-science point of view as a manifestation of the bias-variance error trade-off in the PTM classification model. Large RMSD cutoffs result in many false positives while small RMSD cutoffs result in many false negatives. Minimization of these two sources of error cannot be done independently and the optimal overall score can only be improved by changing the statistical model itself.
   
  \section{DC3 generalization to binary compounds \label{app:binary}}
  The DC3 model for binary compounds was developed using the exact same approach as described above for the single-element DC3 model. The only modifications performed are described in this section.
  
  The most significant modification was with respect to the feature vector. In order to account for the different elements in the A$_n$B$_m$ compounds the feature vectors were augmented such that $\v{x}_i = \left( \v{x}^\text{A}_i, \v{x}^\text{B}_i \right)$, where $\v{x}^\text{X}_i$ is the same feature vector defined for single-element materials in Eq.~\eqref{eq:x} but computed only using those neighbors of atom $i$ that belong to chemical species $X$.
  
  Modifications were also necessary in the outlier detection algorithm because each atom type can now have a different perfect feature vector $\v{x}^*(y_i)$, up to a maximum total of two for the binary structures considered here. The modification occurred during the step where it is necessary to determine whether label $y_i$ is appropriate or $\v{x}_i$ belongs instead to an unknown crystal structure. The distance from $\v{x}_i$ was computed for both perfect feature vectors and the smaller of the two distances was then employed to make the decision. Similarly, the density distribution of $\v{x}^*(y_i)$ was constructed using only the smaller distance from each data point in the synthetic data set to the corresponding perfect feature vectors. Another modification to the outlier detection algorithm is the new value of $\alpha_\text{cutoff} = 2.206$ that optimizes the average accuracy between crystal and liquid for this data set (see Fig.~S5 \cite{Note1}).
  
  Notice also that the InP SNAP potential did not have a stable liquid phase and therefore was not included in the liquid data set for chemically complex structures.
  
  \section{Error bar calculation \label{app:error}}
  The statistical error bar on Table ~\ref{tab:accuracies} and Figs.~\ref{fig:temperature_dependence} and \ref{fig:chemical_complex} represent the $95\%$ confidence interval around the mean computed using the bootstrap method with $200$ samples of the same size as the original distributions.
  
  \section{tSNE calculations \label{app:tSNE}} The plots in Figs.~\ref{fig:tsne}a and \ref{fig:tsne}b were obtained by applying tSNE to a data set composed of $10000$ data points from the MD benchmark data set, $10000$ points from the synthetic training set, and the six perfect feature vectors (one per structure). Figure \ref{fig:tsne}c was obtained by applying tSNE to $10000$ data points corresponding to the hcp structure extracted from the MD data set with temperatures ranging from $0.04\,T_\text{m}$ to $1.16\,T_\text{m}$. For all figures a perplexity value of 1000 was employed.
 
\bibliography{bibliography}
\end{document}